\documentclass[aps,prb,reprint,twocolumn,superscriptaddress,showpacs,floatfix]{revtex4-1}

\usepackage{hyperref}
\usepackage{graphicx}
\usepackage{amsmath}
\usepackage{amssymb}
\usepackage{bbm}
\usepackage{color}
\usepackage{enumitem}
\usepackage{nicefrac}
\usepackage{multirow}
\usepackage{braket}

\newcommand{\beginsupplement}{%
        \setcounter{table}{0}
        \renewcommand{\thetable}{S\arabic{table}}%
        \setcounter{figure}{0}
        \renewcommand{\thefigure}{S\arabic{figure}}%
     }

\allowdisplaybreaks

\usepackage{xfrac}

\begin{document}

\title{Tracking excited states in wave function optimization using density matrices and variational principles}
\author{Lan Nguyen Tran}
\email{lantrann@berkeley.edu}
\affiliation{Department of Chemistry, University of California, Berkeley, California, 94720, USA}
\affiliation{Ho Chi Minh City Institute of Physics, VAST, Ho Chi Minh City, 700000, Vietnam}
\author{Jacqueline A. R. Shea}
\affiliation{Department of Chemistry, University of California, Berkeley, California, 94720, USA}
\author{Eric Neuscamman}
\email{eneuscamman@berkeley.edu}
\affiliation{Department of Chemistry, University of California, Berkeley, California, 94720, USA}
\affiliation{Chemical Sciences Division, Lawrence Berkeley National Laboratory, Berkeley, CA, 94720, USA}

\date{\today}
\begin{abstract}
We present a method for finding individual excited states' energy stationary
points in complete active space self-consistent field theory that is
compatible with standard optimization methods and highly effective
at overcoming difficulties due to root flipping and near-degeneracies.
Inspired by both the maximum overlap method and recent progress in excited
state variational principles, our approach combines these ideas in order to
track individual excited states throughout the orbital optimization process.
In a series of tests involving root flipping, near-degeneracies, charge
transfers, and double excitations, we show that this approach is
more effective for state-specific optimization than either the naive selection
of roots based on energy ordering or a more direct generalization of the
maximum overlap method.
Furthermore, we provide evidence that this state-specific approach
improves the performance of complete active space perturbation theory.
With a simple implementation, a low cost, and compatibility with large active
space methods, the approach is designed to be useful in a wide range of excited
state investigations.
\end{abstract}
\maketitle

\section{Introduction}

While linear response theory has indisputably been the most widely successful paradigm
for making predictions about electronic excitations in chemistry, there remain
important situations in which its use is either difficult or entirely inadvisable.
The most affordable methods in this category, \cite{HeadGordon:2005:tddft_cis}
such as time-dependent density functional theory (TD-DFT) and
configuration interaction singles (CIS),
are well known to face challenges when an excitation significantly alters a
molecule's charge density, as occurs in charge transfer, Rydberg, and core
excitations.
Although TD-DFT has additional concerns, \cite{Dreuw2003LRE}
one major issue in these cases comes from the fact that significant charge density
changes induce orbital relaxation effects that
are not captured in the linear response of a Slater determinant.
\cite{subotnik2011cis_ct}
Even equation of motion coupled cluster theory with singles and doubles excitations
(EOM-CCSD), \cite{Krylov:2008:eom_cc_review}
which is responding around a much more sophisticated wave function,
has difficulty in capturing these relaxation effects in double excitations
\cite{Krylov:2008:eom_cc_review,watts1996eomcc,Neuscamman:2016:var_qmc}
and some core excitations. \cite{Cheng2019eomKedge}
Orbital relaxation effects aside, these linear response methods --- as well as many other
excited state methods including GW theory, \cite{onida2002electronic}
the Bethe-Salpeter equation, \cite{onida2002electronic}
and most variants of the algebraic diagrammatic construction \cite{Dreuw2015algebraic}
--- rely on the assumption that the ground state is single-reference in character.
When this assumption fails, as for example at conical intersections where at least two
states mix strongly, one is typically forced to abandon the realm of weakly correlated
methods altogether.

For decades now, complete active space self-consistent field theory (CASSCF)
\cite{RUEDENBERG:1982:casscf,Werner:1985_1:mcscf,Werner:1985_2:mcscf,ROOS:1987:casscf}
and its generalizations
\cite{olsen1988ras,Malmqvist:2008p3011,Gagliardi2013splitgas}
have been the go-to starting points for
strongly correlated treatments of molecular ground and excited states alike.
Just as Hartree-Fock (HF) theory \cite{Szabo-Ostland} provides a reference state
in which the molecular orbitals are relaxed in the presence of strong
Pauli correlations, CASSCF relaxes the molecular orbitals in the presence of
both Pauli correlations and any other strong correlations that exist within
an active set of orbitals and electrons, typically chosen as a small set
near the Fermi level.
While expanding the size of the active space that can be treated remains
a high priority and has been the focus of much recent work,
\cite{Ghosh2008,Zgid2008dmrgscf,Booth:2015:mcscf_fciqmc,Alavi:2016:fciqmc_casscf,Sharma2017sCICASSCF,tran2017genSEET,tran2018USEET}
an additional important concern in
excited state modeling is the issue of root flipping \cite{diffenderfer1982use,yarkony1998conical,levine2008optimizing,levine2006conical,robb1998potential}. 
In the same way that HF and $\Delta$SCF methods
\cite{bagus1965scf,Pitzer1976scf,argen1991xray,Gill2009dscf}
relax orbitals by finding stationary points on the single-Slater-determinant
energy surface, one would ideally like to prepare a CASSCF reference state
by finding the CASSCF energy stationary point corresponding to the excited
state in question.
However, the common approach of separating the configuration interaction (CI)
coefficient equations and the orbital rotation equations can create problems
when two CI stationary points (typically called roots) cross each
other in energy upon updating the orbitals in pursuit of the (initially)
higher energy root's overall stationary point.
Simple root selection (SRS) --- in which one naively seeks to make
the $n$th root's energy stationary during each orbital rotation step --- will
fail in this scenario, because the $n$th root and the root below it exchange
ordering indefinitely during the ``tick-tock''
cycle of CI and orbital relaxation steps.

Although early methods in the coupled nonlinear optimization of CI coefficients
and orbitals showed promise in bypassing this issue,
\cite{jensen1987direct_casscf}
the most common remedy in use today is the state averaged (SA) approach \cite{werner1981quadratically},
in which one simply minimizes the average energy of multiple states.
The SA approach has the benefit of being compatible with modern, highly efficient tick-tock
optimization schemes and is largely immune to root flipping by virtue of being insensitive
to the ordering of the states within the average.
However, it achieves these advantages by abandoning the quest to locate individual
excited states' stationary points.
This lack of stationarity can be inconvenient when evaluating gradients,
\cite{Martinez2017sacasscf}
but the more troubling issue is the potential loss of accuracy incurred by producing
reference states whose orbitals are not fully relaxed.
Whether this matters in practice of course depends on the system and whether or not
post-CASSCF correlation methods can correct for this shortcoming alongside their
treatment of out-of-active-space correlation effects.
In cases where different states require drastically different orbital relaxations,
such as in the presence of the large dipole shifts associated with charge transfer
excitations \cite{pineda2018excited}, it would be particularly appealing to avoid state averaging
while at the same time avoiding root flipping and retaining compatibility with
tick-tock optimization.

Recently, fully state-specific orbital relaxations have been achieved for weakly
correlated excited states \cite{shea2018} by exploiting simple approximations to
excited state variational principles.
\cite{weinstein1934modified,macdonald1934modified,messmer1969variational,messmer1970variational}
This progress came in the form of a nonlinear optimization in which the
approximated variational principle was minimized under the constraint that the
energy end up stationary with respect to both CI coefficients and orbital rotations.
While one can imagine a straightforward application of these ideas to the
CASSCF wave function, this would lead to a coupled nonlinear optimization
of CI and orbital variables that would not necessarily be cost-competitive
with modern tick-tock optimization schemes.
In this study, we will instead explore whether affordable approximations to
excited state variational principles can aid CASSCF in successfully tracking
individual excited states in the presence of root flipping.
The general idea is similar in spirit to the maximum overlap method (MOM)
\cite{gilbert2008mom,barca2018simple} that works to prevent variational collapse during $\Delta$SCF
optimizations, and indeed we find that state-tracking is most
effective when we combine approximate variational principles with
a MOM-inspired matching condition based on reduced density matrices.
In particular, we demonstrate that this combination is more effective
at finding CASSCF excited state stationary points than either the naive
SRS approach or an approach based on approximate wave function overlaps
as estimated via the CI vectors.
In doing so, we also provide evidence that, although the improvement is sometimes
modest, state specific CASSCF (SS-CASSCF) is a better reference for post-CASSCF methods
than SA-CASSCF.
Finally, as the CASSCF step is rarely the bottleneck when
post-CASSCF methods are in use, our approach does not significantly increase
overall computational cost, and so we are able to recommend its use in general.
In particular, the approach requires only one- and two-body reduced density matrices,
and so should be immediately compatible with the rapidly expanding collection of
large active space methods that have come on to the scene in recent years.
\cite{Ghosh2008,Zgid2008dmrgscf,Booth:2015:mcscf_fciqmc,Alavi:2016:fciqmc_casscf,Sharma2017sCICASSCF,tran2017genSEET}

\section{Theory}
\label{sec:theory}

\subsection{Excited state variational principles}
\label{sec:esvp}

It has long been recognized
\cite{messmer1969variational,messmer1970variational}
that the variational principle
\begin{align}
W = \frac{\braket{\Psi|(\omega-\hat{H})^2|\Psi}}{\braket{\Psi|\Psi}},
\end{align}
has the Hamiltonian eigenstate with energy closest to $\omega$ as its
global minimum.
Recently, this and similar forms involving $\hat{H}^2$ have found
tractable applications in spite of the challenges posed by the squared
Hamiltonian operator, both in single-reference excited state approaches
\cite{vanVoorhis2017sigma,shea2018}
and in quantum Monte Carlo,
\cite{Zhao:2016:dir_tar,zhao2017blocked,Shea:2017:scesvp,
blunt2017charge,robinson2017varmatch,pineda2018excited,
zhao2018gaps,blunt2019charge}
where one can separate the two powers of $\hat{H}$ using a resolution
of the identity
$\sum_I \left|I\right\rangle \left\langle I\right|$
over which statistical sampling may be performed,
\begin{align}
W = \frac{1}{\braket{\Psi|\Psi}}
    \sum_I{\braket{\Psi|(\omega-\hat{H})|I}\braket{I|(\omega-\hat{H})|\Psi}}.
\end{align}
In the context of CASSCF, one can simplify this resolution of the
identity significantly by exploiting the fact that the space of wave
functions that $\hat{H}$ can connect the CASSCF reference to is
exactly spanned by the active space itself and all internally contracted
singles and doubles excitations out of it.
For ease of discussion, we will consider the case in which there
are no closed shell orbitals, noting that the generalization to the
full theory with a closed shell is straightforward.
With active spin orbitals labeled by $i$ and $j$ and virtual
spin orbitals labeled by $a$ and $b$ and the trivial assumption that
the CASSCF wave function $\Phi$ is normalized, the resolution of the
identity form of the variational principle can thus be organized as 
\begin{align}
W &= W_0 + W_1 + W_2, \\
W_0 &= (\omega-E)^2, \\
W_1 &= \sum_{ia}\left|\braket{\Phi|
            \hspace{.5mm} \hat{H}
            \hspace{.5mm} \hat{a}_a^{+}
            \hspace{.5mm} \hat{a}_i
            \hspace{.5mm} 
            |{\Phi}}\right|^2, \\
W_2 &= \sum_{ijab}\left|\braket{\Phi|
            \hspace{.5mm} \hat{H}
            \hspace{.5mm} \hat{a}_a^{+}
            \hspace{.5mm} \hat{a}_b^{+}
            \hspace{.5mm} \hat{a}_j
            \hspace{.5mm} \hat{a}_i
            \hspace{.5mm} 
            |{\Phi}}\right|^2.
\end{align}
Of particular note is that, by virtue of being limited to
internally contracted single and double excitations, a full
evaluation of $W$, should one wish to pursue it, should be
similar in complexity to constructing the right hand side of
the first order wave function equation in complete active
space second order perturbation theory (CASPT2).
\cite{andersson1990caspt2,andersson1992caspt2}

Inspecting the three components of $W$, we find that they have
simple interpretations if we take $\Phi$ to be
a root of the complete active space CI problem.
First, $W_0$ is simply stating that the eigenstate we are after should have an
energy close to $\omega$.
Second, noting that CI roots' energies are already stationary with respect to
the CI coefficients, we see that $W_1$ is simply a measure of how close
the wave function is to being an overall energy stationary point
as each of its terms is proportional to the energy derivative with respect
to an active-to-virtual orbital rotation.
Indeed, if $\Phi$ is a CI root and $W_1=0$, then
the CASSCF energy is stationary.
Finally, noting that $W_1$ and $W_2$ are unaffected if we make the
replacement $\hat{H}\rightarrow\hat{H}-E$, we see
that $W_2$ contains all the terms that would need to be zero
in addition to those in $W_1$ in order for the energy variance
$\sigma^2=\braket{(\hat{H}-E)^2}$ to be zero.
As an exact eigenstate of $\hat{H}$ will be both an energy stationary
point and a zero variance state, we see that, together, $W_0$, $W_1$,
and $W_2$ are simply a least-squares way of saying that we want
the CASSCF wave function that is closest to the exact energy
eigenstate near $\omega$.

In principle, we could use these expressions to follow the approach of excited
state mean field (ESMF) theory \cite{shea2018} and minimize the Lagrangian
\begin{align}
    \label{eqn:esmfL}
    L = W - \vec{\mu} \cdot \frac{\partial E}{\partial \vec{\nu}},
\end{align}
with respect to a variable set $\vec{\nu}$ that contained both
orbital rotations and the CI coefficients.
This approach uses the variational principle $W$, or an approximatin
to it, to guide an optimization to the desired energy stationary point
via constrained Lagrangian minimization.
However, making such an approach cost-competitive with CASSCF tick-tock
optimization methods would not be trivial, and in this study we
seek to exploit approximations to $W$ in a much simpler context.

Within the standard tick-tock approach of switching back and forth
between Davidson CI diagonalizations and orbital optimization steps,
it is typically the CI step that dominates the cost when the active
space gets large.
We will therefore leave the Davidson step unchanged for now and consider
using approximations to $W$ in the remainder of the optimization.
In our initial testing, we have found that while an L-BFGS minimization
of $|\nabla L|^2$ with respect to orbital rotation variables
is effective when we set $W\approx W_0 + W_1$, it is in most
cases equally effective to relax the orbitals by a simple L-BFGS
minimization of $|\nabla E|^2$.
This observation suggests that simple generalizations of
standard Newton-Raphson style orbital optimizations in which the
line search is set to minimize $|\nabla E|^2$ rather than $E$ itself
are likely adequate in many cases
(and much faster than $L$-based quasi-Newton methods),
even if they lack the strong resistance to ground state
collapse offered by $L$.
However, orbital relaxation for a particular CI root is only part
of the process of converging to a desired stationary point
during a CASSCF tick-tock optimization.
The method used for selecting which CI root to relax the orbitals
for is equally important and in general quite challenging.
As we now discuss, it is in this area that
we find excited state variational ideas to be most helpful.

\subsection{Maximum overlap analogues}
\label{sec:mom}

First introduced by Gill and coworkers, \cite{gilbert2008mom}
the maximum overlap method (MOM) helps prevent variational
collapse to the ground state when attempting to locate
excited state solutions to the Roothan equations in
Hartree Fock or density functional theory.
The idea is to choose the excited state orbital occupation
for the the orbitals generated by a newly diagonalized Fock
matrix by selecting the orbitals that give the largest
combined overlap with the molecular orbitals from the
previous iteration of the method, or some target set of
molecular orbitals believed to be similar to those of the
desired excited state.
In essence, MOM is an attempt to follow the trail of a
particular excited state through the sequence of discreet
(and sometimes large) orbital relaxations that occur
over the course of the self consistent field iterations.
This goal is very similar to what we desire when faced with
a root flipping problem in CASSCF:  we wish to track a
particular excited state through the sequence of discreet
(and sometimes large) changes to the Davidson CI roots
that occur over the course of a tick-tock CASSCF optimization.

If one wished to pursue a strategy similar in spirit to that
of the orbital-overlap-based MOM, a simple strategy would be
to hope that changes to the CI vector $\vec{c}$ between
iterations were never too large and simply define a tracking
function $Q_{\mathrm{MOM}}$ based on the approximate wave
function overlap
\begin{align}
Q_{\mathrm{MOM}}(\vec{c}\hspace{0.5mm})
 = \vec{c}_t \cdot \vec{c},
\end{align}
between each of the current iteration's CI roots and some
target CI vector $\vec{c}_t$, taken here to be the CI
vector selected in the previous iteration.
Of course, like MOM itself, this strategy, and any
tracking strategy based on measuring wave function
similarities across iterations, will
not necessarily be robust in cases where the iterative
method makes large changes to the wave function in a
single iteration.
Ideally, we would therefore like to augment this strategy
with a component that is less dependent on
iteration-to-iteration similarity.
The central finding of this study is that, in combination
with a measure of similarity that
is more robust than CI vector dot products, the excited
state variational principle $W$ can help in this way.

In hopes of creating a more robust state-tracking function,
we define the following quality measure for a newly generated
Davidson root with CI vector $\vec{c}$. 
\begin{align}
\label{eqn:qwg}
Q_{W\Gamma}(\vec{c}\hspace{0.5mm}) &=
   W_0(\vec{c}\hspace{0.5mm})
 + W_1(\vec{c}\hspace{0.5mm})
 + D(\vec{c}\hspace{0.5mm}), \\
\label{eqn:ddef}
D(\vec{c}\hspace{0.5mm}) &=
 \frac{\hspace{1mm} || \hspace{.5mm} 
       \Gamma_t - \Gamma(\vec{c}\hspace{0.5mm})
       \hspace{.5mm} || \hspace{1mm}}
       {n_{{}_{\mathrm{CAS}}}}.
\end{align}
Here $\Gamma_t$ and $\Gamma(\vec{c}\hspace{0.5mm})$ are the
one-body reduced density matrices (1RDMs) for the target
wave function and the current root in question, respectively,
with $\Gamma_t$ rotated into the current orbital basis
in order to reduce sensitivity to orbital changes. 
While one would like to do something similar for
$Q_{\mathrm{MOM}}$, it is not obvious how this would be done
as active-virtual orbital rotations prevent $\vec{c}_t$ from
being expressible within the new orbitals' active space.
Note that we divide the Frobenius norm of the
1RDM difference by the number of active orbitals
$n_{{}_{\mathrm{CAS}}}$ in order to make the relative
importance of $W_0$, $W_1$, and $D$ less sensitive to the
size of the chosen active space.
This choice is built on the idea that the
out-of-active-space parts of the density matrix difference
are small, as the occupation vector is fixed for these
orbitals and we do not expect qualitative changes in
the closed-shell orbital shapes.
While it may be that a different balance between $W_0$, $W_1$,
and the density matrix difference is optimal, in practice
we have found that this choice is effective in most cases.
While one could of course also include $W_2$ in this quality measure,
we have chosen to omit it due to its relatively high cost of
evaluation and our observation that $Q_{W\Gamma}$
is quite effective even without it.

As for why we chose a density matrix difference for our measure of
iteration-to-iteration similarity, the logic is that we could have
measured similarity via a combination of any number of wave function
properties (e.g.\ dipole moment), but a large number of properties
are themselves determined via the 1RDM.
Of course, one could also consider 2RDM differences, but for
simplicity's sake we have for now limited our investigation
to differences of 1RDMs.
As our results will demonstrate, the quality measure $Q_{W\Gamma}$,
although not perfect and certainly less sophisticated than it could
be, is far more effective at dealing with root
flipping when attempting to track a specific excited state through
a CASSCF tick-tock optimization than either SRS or $Q_{\mathrm{MOM}}$.

\subsection{Optimization procedure}
\label{sec:op}

To summarize, the overall optimization procedure that we test
here involves the following steps.

\begin{enumerate}
    \item Choose an orbital basis as an initial guess, perform
    an initial CASCI calculation, and select the root
    that will be targeted for state-specific convergence. 
    \item \label{step2} Relax the orbital coefficient matrix
    $\bm{C}=\bm{C}_0 e^{\bm{X}}$ via an L-BFGS minimization of
    either $|\nabla_{\bm{X}} L|^2$ + $|\nabla_{\bm{\mu}} L|^2$
    (with $W\approx W_0 + W_1$) or the even simpler objective function
    $|\nabla_{\bm{X}} E|^2$.
    While the latter is in principle more prone to variational
    collapse, we find that in the cases studied here the two
    orbital rotation objective functions lead to the same results.
    Either way, we implement the gradients needed for L-BFGS within
    the TensorFlow automatic differentiation framework.
    \cite{tensorflow}
    \item Perform a new CASCI calculation
    (via the pySCF package \cite{pyscf2018})
    to obtain the low-lying roots of the appropriate
    space and spin symmetry in
    new orbital basis.
    \item Select the root with the lowest value for
    $Q_{\mathrm{MOM}}$ or $Q_{W\Gamma}$, depending on
    which quality measure is being employed.
    If instead one is following the SRS approach, then
    simply select the root based on
    its position in the energy ordering.
    \item Return to step~\ref{step2} and continue
    until an overall CASSCF energy stationary point is found.
\end{enumerate}

\section{Results and discussion}

We now turn to a collection of molecular examples
with which we seek to gain insight into three key questions.
First, how effective is the $W\Gamma$ approach at overcoming
root flipping in comparison to the MOM and SRS methods?
Second, can the $W\Gamma$ approach succeed in cases where
there are nearly degenerate states that cannot
be distinguished by $W$ alone?
Finally, do the SS-CASSCF
solutions that this approach helps us find
outperform their SA-CASSCF
counterparts as reference functions for
post-CASSCF methods like CASPT2 and Davidson-corrected
multi-reference configuration interaction with
singles and doubles (MRCI+Q)?

These questions are studied in the molecular systems
summarized in Table \ref{tab:cas}.
The cc-pVDZ basis set was used throughout, as were
CASSCF energy, orbital gradient, and density matrix
convergence thresholds of $10^{-7}$, $10^{-4}$,
and $10^{-4}$, respectively.
In most cases, we began our optimizations in the HF orbital basis,
but in MgO we began in the LDA basis instead because
both the MOM and $W\Gamma$ approaches converged more rapidly
from an LDA starting point.
All CASPT2 and MRCI+Q calculations were carried out
using Molpro, \cite{MOLPRO_paper}
while EOM-CCSD calculations were performed using QChem.
\cite{shao:2015:qchem}
Note that all post-CASSCF methods employed the usual
frozen-core approximation, but for CASSCF itself
the core was frozen or not as described in Table \ref{tab:cas}.
For MRCI+Q, we used Molpro's default convergence thresholds
for the energy and density matrix in all systems except for
MgO, where we found it necessary to set them both to
$10^{-5}$ in order to avoid unstable oscillations in
the energies.

\begin{table}[!h]
  \normalsize
  \caption{\label{tab:cas} \normalsize Summary of frozen and active orbitals used for all systems.}
  \begin{tabular}{llllcccccc}
    \hline \hline		
    &Molecules &frozen orbitals &active orbitals \\ 
    \hline
    &LiH	  & none &4e,4o: Li $1s2s2p_z$ H $1s$ \\
    &O$_3$    & O $1s2s$ &12e,9o: O $2p$ \\
    &CH$_2$O  & C $1s2s$ O $1s2s$&8e,8o:  C $2p$ O $2p$ H $1s$\\
    &MgO      & none&8e,8o:	Mg $3s3p$ O $2s2p$ \\
    \hline \hline
  \end{tabular}
\end{table}


\subsection{LiH}
\label{sec:lih}

\begin{figure*} [!t]\centering
  \includegraphics[width=16cm]{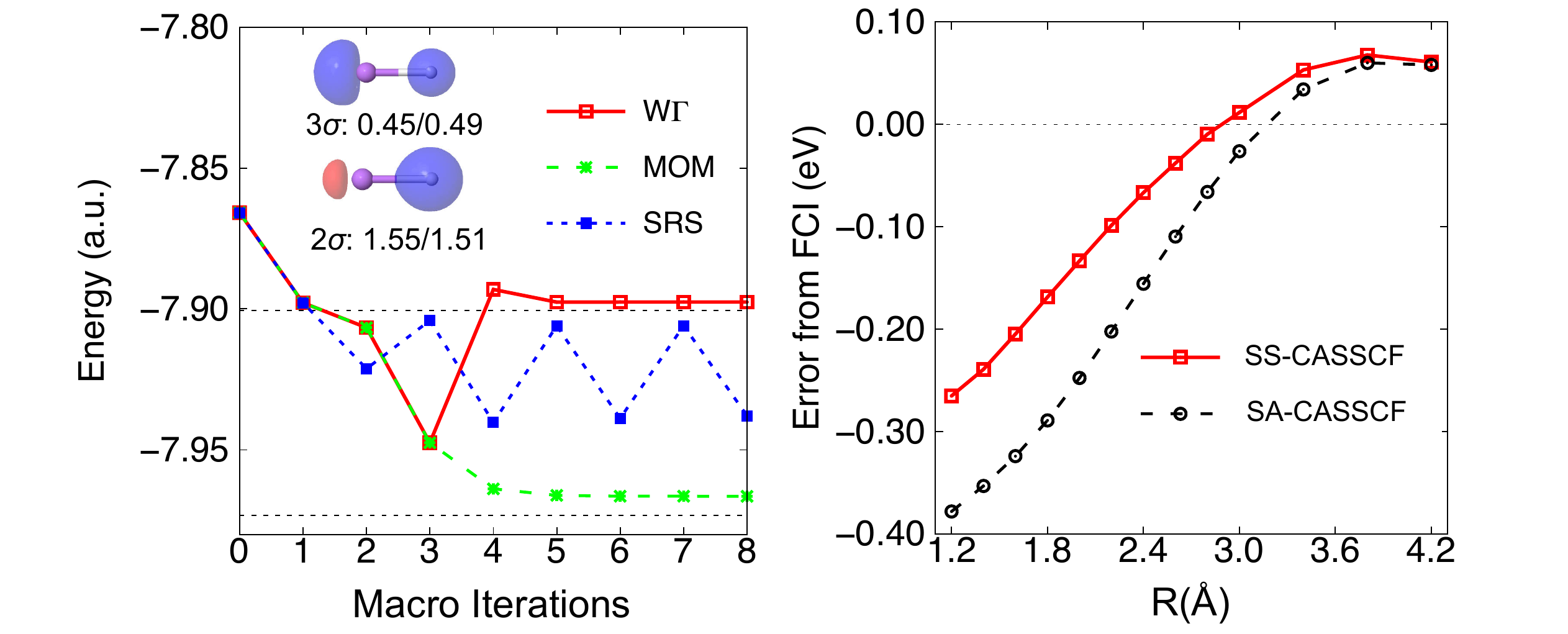}
  \caption{\normalsize Left panel: The change in energy during CASSCF
  optimizations of the first excited state ($A^1\Sigma^+$)
  of LiH at a bond distance of 2.6 \r{A}.
  The SRS, MOM, and $W\Gamma$ approaches all start with the Hartree-Fock
  orbitals as the initial guess.
  For reference, the horizontal lines show the FCI energies for
  the ground and excited states.
  The inset shows the $2\sigma$ and $3\sigma$ natural orbitals
  and occupation numbers for the excited state
  as produced by the $W\Gamma$ approach.
  The occupation numbers from FCI (the second number) are given for comparison. Li and H atoms are in purple and white, respectively.
  Right panel: Excitation energy errors relative to FCI
  at different bond distances $R$.}
  \label{fig:LiH}
\end{figure*}

Let us begin with the well-known example of root
flipping that occurs in LiH.
Near the equilibrium bond length ($\sim$1.6\r{A}),
this molecule's ground state ($X^1\Sigma^+$) is basically ionic,
while the first excited state ($A^1\Sigma^+$)
is predominantly neutral due to a charge transfer excitation.
As the bond length is stretched, however, the ground state
becomes increasingly neutral and the excited state
increasingly ionic.
At intermediate distances, the system shows an avoided crossing
between these states that causes a well-recognized example
of root flipping for SRS, as seen in the left panel of
Figure \ref{fig:LiH}.
As the orbitals are optimized for the excited state, the
energy of the ground state CASCI root rises while that
of the excited state root falls.
Soon the two roots flip in the energy ordering, at which point
SRS is now effectively trying to optimize the orbitals for
the ground state, which in turn causes the energy ordering
to flip back.
This process continues indefinitely, preventing the SRS approach
from converging at all.

After the first orbital relaxation, both the MOM and
$W\Gamma$ approaches recognize that the ordering of
the roots has changed and select the lower root,
thus diverging from the optimization path of SRS.
Their two quality measures then select the same root
for one more iteration, at which point orbital 
relaxations clearly work to push the state towards
the ground state, as can be seen by the large energy
lowering between macro iterations 2 and 3.
It is here that the two methods diverge, with MOM's
quality measure selecting the root that ultimately
becomes the ground state.
The $W\Gamma$ measure, on the other hand,
successfully keeps track of the excited state root and
ultimately converges to a CASSCF stationary point that
clearly corresponds to the desired excited state.
As seen in the right panel of Figure \ref{fig:LiH}
and in the more detailed tabulation within the
Supporting Information (SI),
this success is repeated at all bond distances, and
we find that the excitation energies based on SS-CASSCF
energy differences tend to be a bit more accurate than
those based on equally-weighted SA-CASSCF energies.

\subsection{Asymmetric O$_3$}
\label{sec:o3}

\begin{figure*} [!t]\centering
  \includegraphics[width=14.0cm]{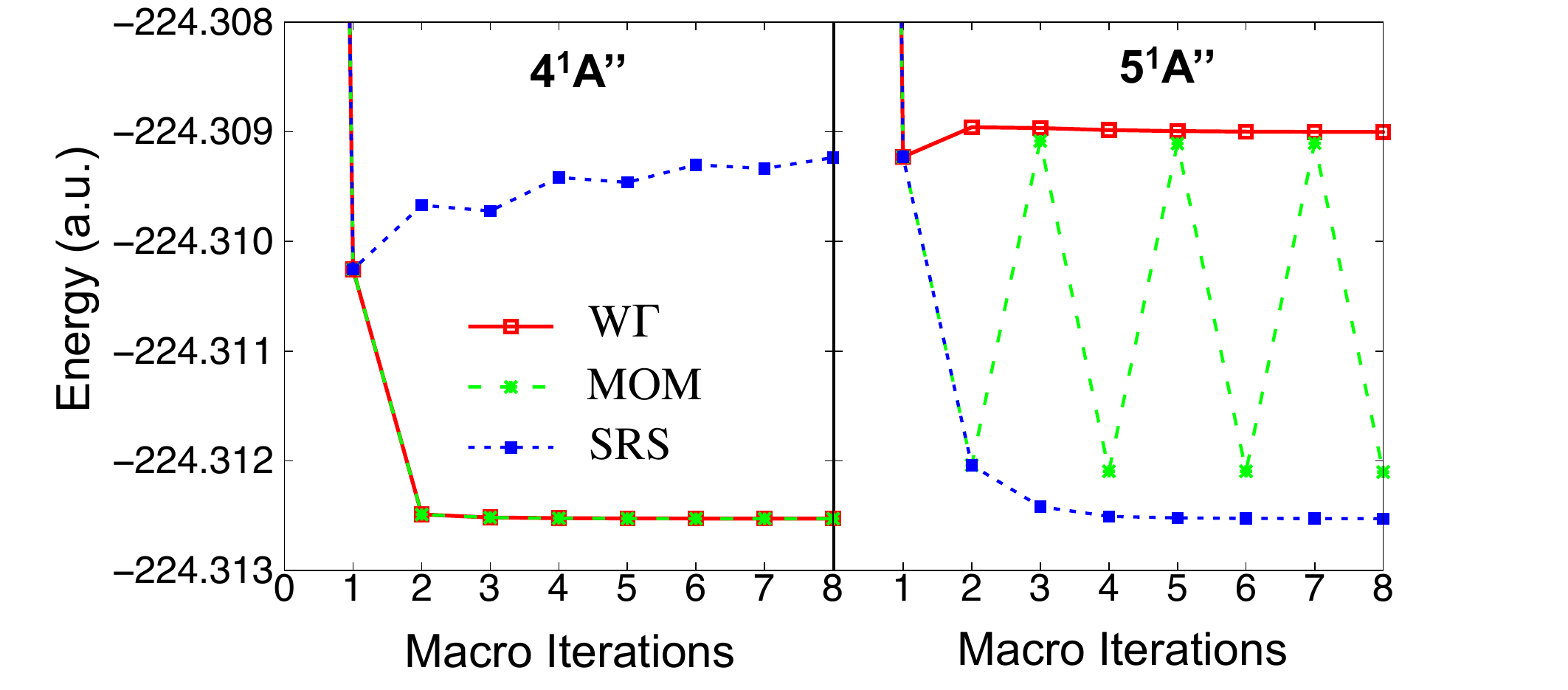}
  \caption{\normalsize
  The energy of the selected root after each CASCI
  calculation during SS-CASSCF optimizations
  of asymmetric O$_3$.
  The left and right panels show results when the
  SRS, MOM, and $W\Gamma$ methods are initialized with the
  $4^1A''$ and $5^1A''$ CASCI roots, respectively,
  in the initial Hartree-Fock orbital basis.
  }
  \label{fig:O3_iter}
\end{figure*}

We now turn to asymmetrically stretched ozone, in which
we focus our attention on two nearly degenerate excitations
at the somewhat arbitrary asymmetric geometry
$R_{\mbox{\scriptsize O$_1$}\mbox{\scriptsize O$_2$}}=1.3$ \r{A},
$R_{\mbox{\scriptsize O$_2$}\mbox{\scriptsize O$_3$}}=1.8$ \r{A},
$\angle \hspace{.5mm} \mathrm{O_1 O_2 O_3} = 120^{\circ}$.
These states correspond to the fourth and fifth SS-CASSCF
excitation energies (relative to the ground $1^1A'$ state)
in this $C_s$ geometry's $A''$ representation, and so
we will label them as $4^1A''$ and $5^1A''$
even though we will see that when out-of-active-space
correlation is included it is no longer
obvious which actually has the lower energy.
The $4^1A''$ state's primary component is the
configuration with four singly occupied orbitals
resulting from a
double excitation that moves one electron each
from the two highest-occupied $A'$ orbitals into
the lowest unoccupied $A'$ and $A''$ orbitals,
whereas the $5^1A''$ state's primary component is the
single excitation from the highest-occupied $A''$
orbital into the lowest unoccupied $A'$ orbital.
That said, both of these states have nontrivial coefficients
on the other state's primary component, meaning that
neither state can be described as a simple single excitation.
To add to the confusion, these states show up in the
reverse order ($5^1A''$ before $4^1A''$) in the
energy-ordered CASCI roots in the initial HF
orbital basis, and so root flipping is essentially
guaranteed during state-specific optimization.

Unlike the lowest three $A''$ states ---
all of which converge without trouble using SRS ---
this pair of states proves a nontrivial challenge for
state specific optimization.
As shown in Figure \ref{fig:O3_iter}, the MOM
finds the $4^1A''$ stationary point without difficulty
when starting from the corresponding (fifth) $A''$
root of the HF-orbital CASCI, but becomes
trapped in a limit cycle when we target the
$5^1A''$ by starting from the corresponding (fourth)
root.
SRS succeeds in finding both stationary points,
but it does so by making a root-flipping-induced
qualitative swap in the state being tracked.
This (or perhaps collapse to a lower state) is to be
expected, as the energy ordering of these states
is reversed at their individual stationary points
as compared to the initial HF-orbital CASCI.
The practical results are twofold.
First, the state the user gets is not the same
excited state as was initially targeted.
Second, the convergence of the $4^1A''$ state
proves to be very slow with the gradient still
not quite converged even after 
100 macro iterations.

\begin{figure*} [!t]\centering
  \includegraphics[width=14.0cm]{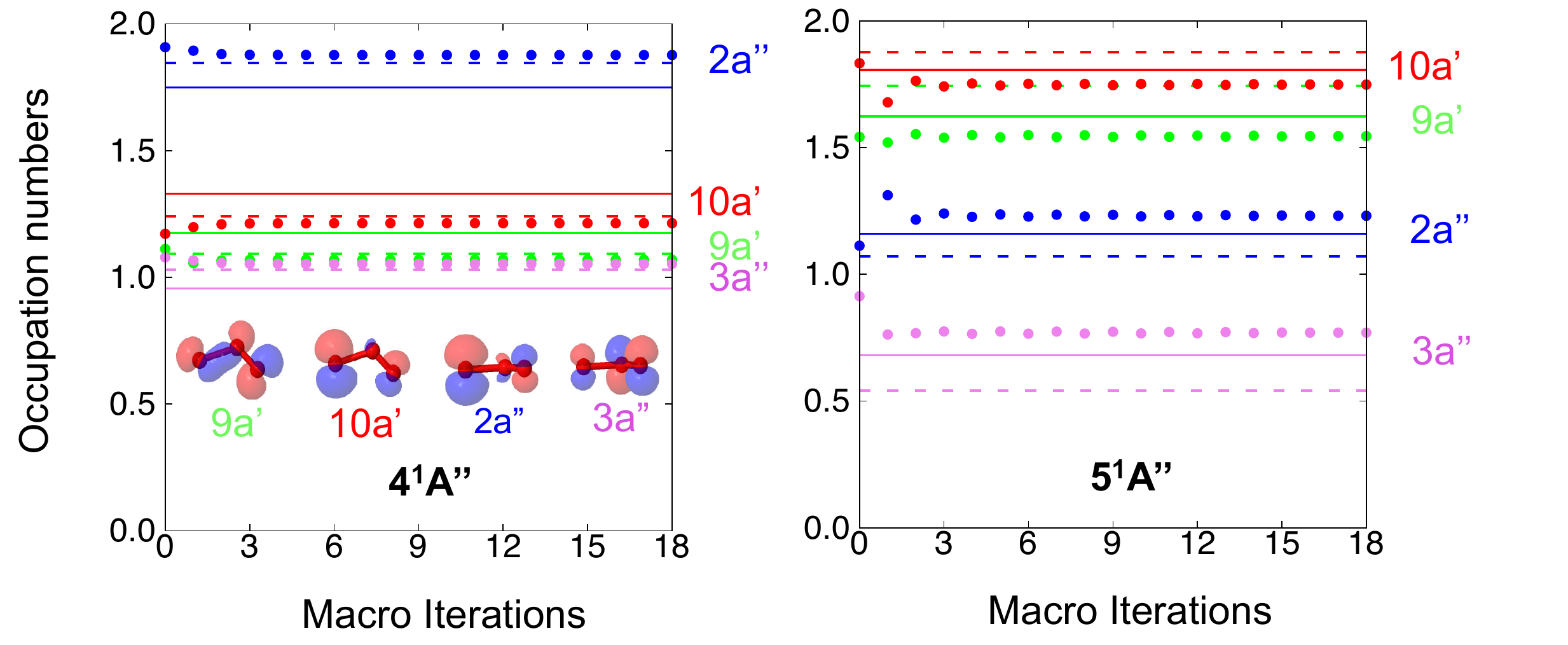}
  \caption{\normalsize
  The natural orbital occupation numbers after each CASCI
  calculation during the $W\Gamma$-guided CASSCF optimizations
  of the $4^1A''$ (left panel) and $5^1A''$ (right panel)
  states in asymmetric O$_3$, with the $0$th iteration
  corresponding to the initial HF-orbital CASCI.
  For comparison, the solid and dashed horizontal lines
  give the corresponding final occupation numbers from
  3-state and 6-state SA-CASSCF calculations, respectively.
  The inset shows the final shapes of these four natural
  orbitals for the SS-CASSCF $4^1A''$ case.
  Note that the shapes for the SS-CASSCF $5^1A''$ natural orbitals look
  very similar even though the natural occupation numbers largely differ.}
  \label{fig:O3_occ}
\end{figure*}

The $W\Gamma$ approach, in contrast, converges rapidly for
both the $4^1A''$ and $5^1A''$ states when it
is initiated with
the corresponding root from the initial HF-orbital CASCI.
Unlike SRS, these optimizations do not involve any
switching between qualitatively different states,
despite the fact that the energy ordering does change.
We explicitly verify that the desired state is
tracked at every iteration of the optimization
in Figure \ref{fig:O3_occ},
which shows both how little the relevant active
space natural orbital occupation numbers change
as well as how clearly qualitatively different
the $4^1A''$ and $5^1A''$ states are.
At convergence, we see that the stationary points found in
the $W\Gamma$ approach correspond closely to both the initial
CASCI wave functions and the results from equal-weighted
3-state ($1^1A'$, $4^1A''$, $5^1A''$) and
6-state ($1^1A'$, $1^1A''$, $2^1A''$, $3^1A''$, $4^1A''$, $5^1A''$)
SA-CASSCF calculations.

\begin{figure} [!h]\centering
  \includegraphics[width=8.0cm]{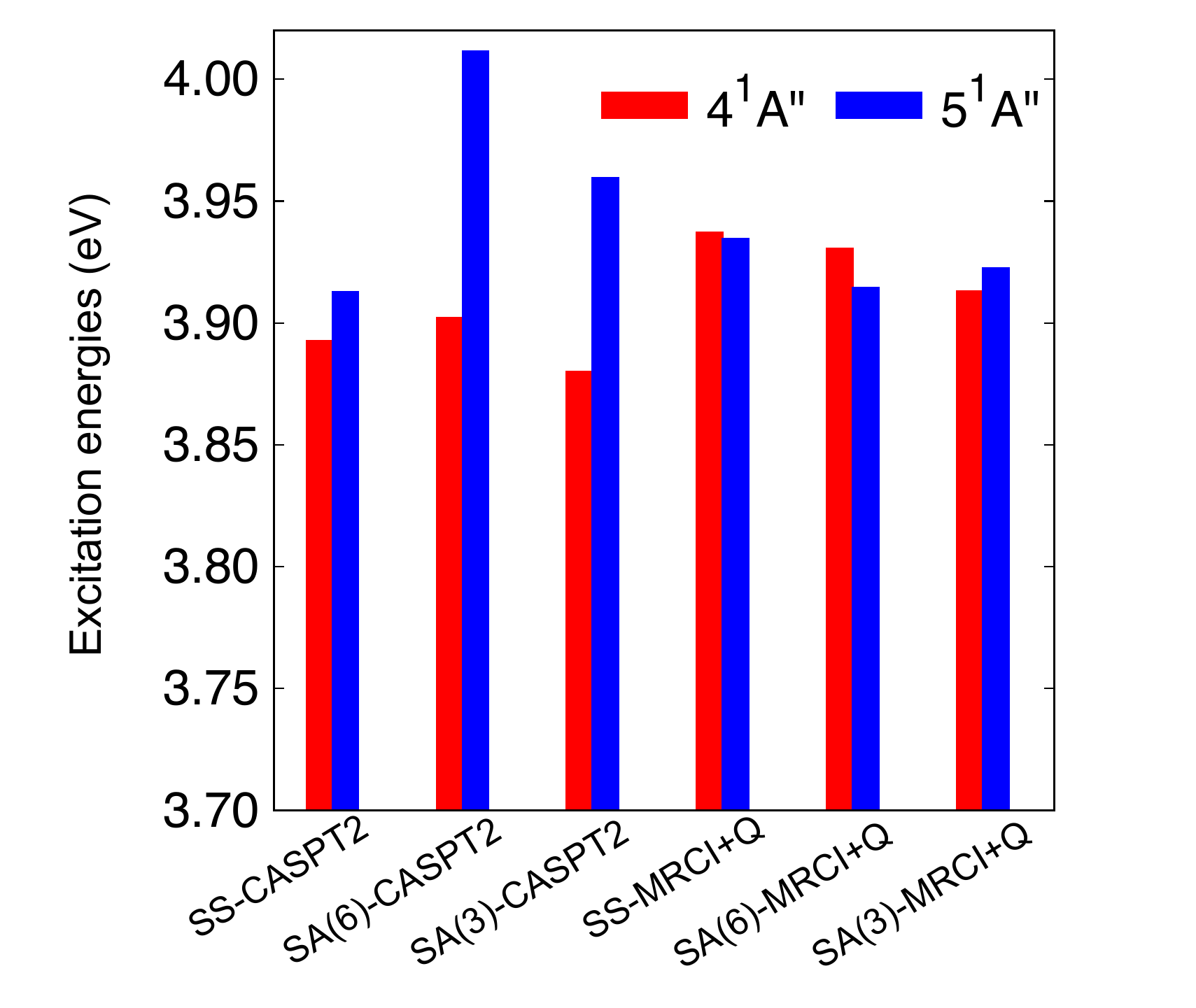}
  \caption{\normalsize
  CASPT2 (with no level shift) and MRCI+Q excitation energies
  for the $4^1A''$ and $5^1A''$ states of asymmetric O$_3$
  when starting from SS-CASSCF and 3- or 6-state SA-CASSCF
  wave functions.
  }
  \label{fig:O3_gaps}
\end{figure}

Of course, when comparing to SA-CASSCF, the more pressing
question is whether the ability to find the correct
stationary points in SS-CASSCF offers
any advantages in final accuracy.
To investigate this question, we have performed both
CASPT2 and MRCI+Q calculations based on the different
SS and SA wave functions, the results of which are
displayed in Figure \ref{fig:O3_gaps}.
We find that MRCI+Q predicts the states to be
nearly degenerate with excitation energies within
0.02 eV of each other regardless of whether we
base it on either of the SA-CASSCF calculations' orbitals
or on the SS-CASSCF orbitals.

Note that the latter SS-MRCI+Q case amounts to three
different MRCI+Q calculations, one each in the ground,
$4^1A''$, and $5^1A''$ state's orbital basis, with the
appropriate MRCI+Q root's energy extracted in each case.
CASPT2, on the other hand, only predicts the degeneracy
to be within 0.02 eV when based on the SS-CASSCF
wave functions, showing gaps of more than 0.07 eV
in both the 3-state and 6-state SA cases
(note that we used single-state zero order Hamiltonians
for CASPT2 both when starting from SA-CASSCF and SS-CASSCF).
One way to explain these findings is to note that the
out-of-active-space single excitations that provide
state-specific orbital relaxations are treated
perturbatively in CASPT2, and so the fact that SS-CASSCF
has already provided state-specific orbitals puts us
in a regime where the perturbative assumption of small
singles (and doubles) coefficients is more likely to
be satisfied in practice.
Although the improvement is modest, the SS-CASPT2 is
in closer agreement with MRCI+Q than is SA-CASPT2,
which is encouraging as it suggests that SS-CASSCF may
help the lower-cost CASPT2 method agree better with the
higher-cost and typically higher-accuracy MRCI+Q method.
As we will see, this same pattern plays out again and again
in the states studied below.

\subsection{CH$_2$O}
\label{sec:ch2o}

\begin{figure*} [!t]\centering
  \includegraphics[width=12.0cm]{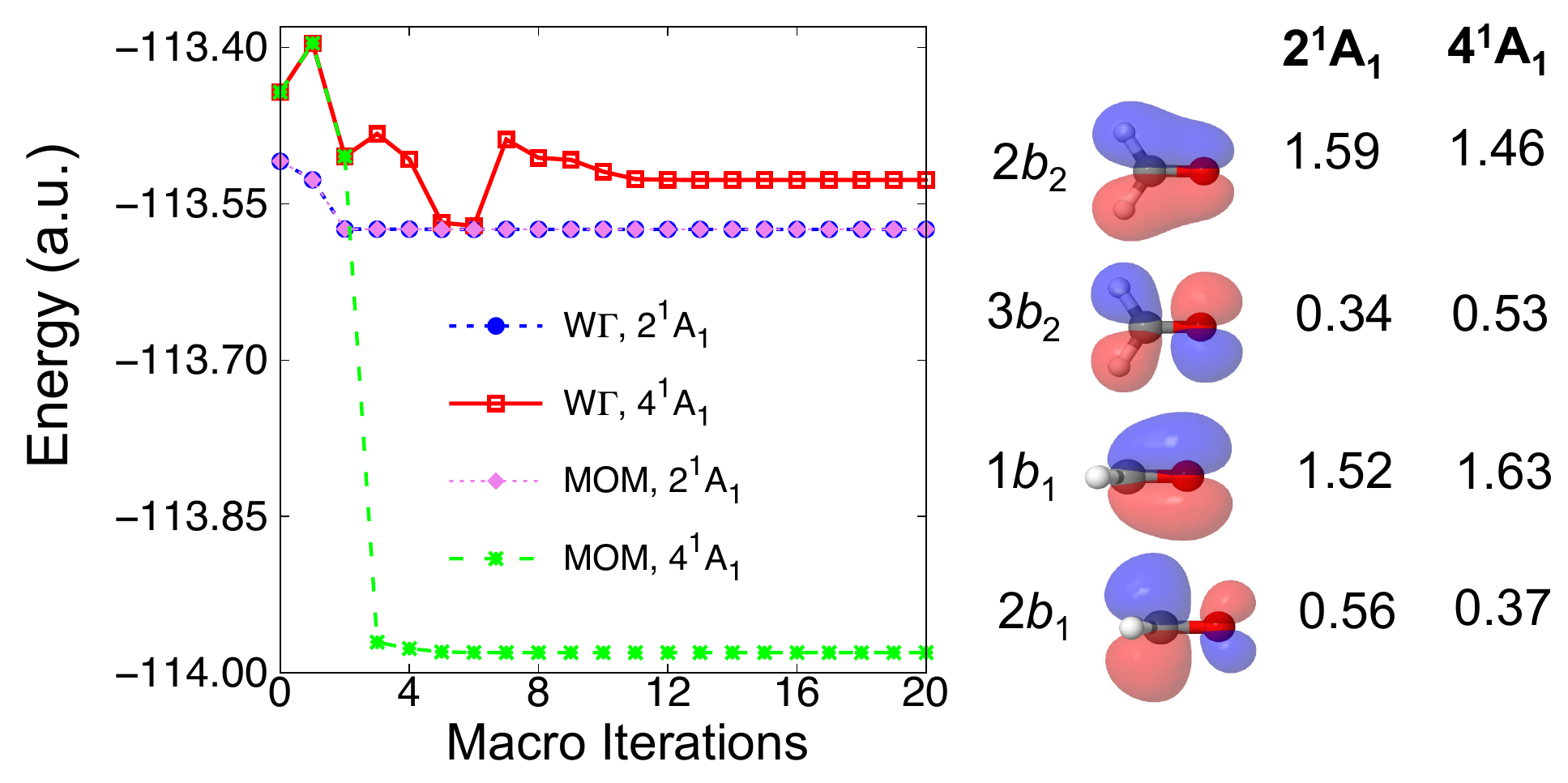}
  \caption{\normalsize
  Left panel: changes in the energy
  during MOM and reweighted $W\Gamma$ CASSCF optimizations of
  two CH$_2$O excited states.
  Right panel: plots of the four most important orbitals
  contributing to these states (we are showing
  SS-CASSCF natural orbitals for $4^1A_1$, the corresponding
  $2^1A_1$ orbitals are quite similar)
  as well as the SS-CASSCF natural orbital occupation numbers.
  }
 \label{fig:H2CO}
\end{figure*}

We now turn to formaldehyde --- with $C_{2V}$ geometry
$R_{\mbox{\scriptsize C}\mbox{\scriptsize O}}=1.2 \r{A},
R_{\mbox{\scriptsize C}\mbox{\scriptsize H}}=1.1 \r{A},
\angle(\mbox{H,C,H})=116.43^{\circ}$
--- to investigate a case of two excited states
with very similar components.
According to EOM-CCSD, there are two excited states
in the $^1A_1$ symmetry sector whose principal components
are superpositions of the $1b_1 \rightarrow 2b_1$ and
$2b_2 \rightarrow 3b_2$ single-electron transitions
(orbitals are plotted in Figure \ref{fig:H2CO}).
These two states, which
in the HF-orbital CASCI are the
$2^1A_1$ and $4^1A_1$ states, are dominated by
the sum of or difference between these components
and turn out to have very similar natural orbital
occupation numbers (see Figure \ref{fig:H2CO}).
One might worry that such similarity could
confuse a density matrix based approach,
whereas it
looks at first glance like the MOM approach should
be effective as the CI vectors are clearly quite
different, at least according to EOM-CCSD.
As we will see, however, the more difficult of
these two states confounds both the MOM
and $W\Gamma$ approaches as we defined them
in Section \ref{sec:theory},
and we were only able to succeed with SS-CASSCF
by \textit{increasing}
the weighting of the density matrix difference
within our $Q_{W\Gamma}$ measure.

Starting all CASSCF optimizations with HF orbitals
as the initial guess, we find that the SRS, MOM,
and $W\Gamma$ approaches all succeed at finding the
energy stationary point associated with the $2^1A_1$ state.
However, even though we are using a nearly full-valence
active space, none of these approaches succeeded
for the $4^1A_1$ state when applied as described in
the theory section.
For example, Figure \ref{fig:H2CO} shows how the MOM approach
collapses to the ground state after just a few macro iterations.
Interestingly, this $4^1A_1$ state proves to be an example of
a case in which the relative weighting of the $W_1$, $W_2$, and $D$
components matters when employing the $W\Gamma$ quality measure of
Eq.\ (\ref{eqn:qwg}).
When we double the weighting of the density matrix
difference, i.e.\ $Q_{W\Gamma}\rightarrow W_0 + W_1 + 2D$,
the $W\Gamma$ approach converges successfully
to the $4^1A_1$ state, as seen in Figure \ref{fig:H2CO}.
Thus, despite this being a case where it is not obvious 
a priori that root flipping will occur and despite the
states in question having easily distinguished CI vectors
at the outset of the optimization,
we find that the approach that incorporates density matrix
differences is more effective at converging to the relevant
stationary points than either the SRS or MOM approaches.

\begin{table}[!b]
  \normalsize
  \caption{\label{tab:H2CO} \normalsize
  Excitation energies (eV) for the $2^1A_1$ and $4^1A_1$ states
  of CH$_2$O along with the IPEA level shifts (a.u.) used for CASPT2.}
  \begin{tabular}{cccccccccc}
    \hline \hline		
    &Method &Shift	&$2^1A_1$ &$4^1A_1$ \\ 
    \hline	
    &SA(w1)-CASPT2  &0.1 &9.71  &10.94  \\ 
    &SA(w1)-CASPT2  &0.2 &9.83  &11.15  \\ 
    &SA(w1)-MRCI+Q  &N/A &9.86  &11.10  \\ 
    \hline
    &SA(w2)-CASPT2  &0.1 &N/A &11.03  \\ 
    &SA(w2)-CASPT2  &0.2 &N/A &11.19	 \\ 
    &SA(w2)-MRCI+Q  &N/A &N/A &11.21  \\ 
    \hline
    &SS-CASPT2	    &0.1 &9.74  &11.24  \\ 
    &SS-CASPT2	    &0.2 &9.83  &11.35	 \\ 
    &SS-MRCI+Q	    &N/A &9.83  &11.30  \\ 
    \hline
    &EOM-CCSD	    &N/A &10.08  &11.38	 \\ 
    \hline \hline
  \end{tabular}
\end{table}

To ascertain how much the ability to find these states' stationary
points matters at the end of the day, we compare CASPT2 and MRCI+Q
results based on SS- and SA-CASSCF in Table \ref{tab:H2CO}.
Here we have employed two SA schemes, one that gives equal weighting
to the first four $^1A_1$ states (w1) and one that uses weightings of
$(0.6,0.1,0.1,0.2)$ and $(0.2,0.1,0.1,0.6)$ when modeling the ground and
$4^1A_1$ states, respectively (w2).
As CASPT2 was found to suffer from intruder states, we have employed
two different level shifts in each case.
One difference that is noticed immediately is that, compared to
SA-CASPT2, the excitation energies for SS-CASPT2 are less sensitive
to changes in the level shift, which is certainly a desirable property.
For the $2^1A_1$ state, we see that CASPT2 and MRCI+Q excitation
energies are not so sensitive to the choice between SA-CASSCF and
SS-CASSCF.
Note that for this state, there is some question about how accurate EOM-CCSD is expected to be,
as the CASSCF CI vectors all show 
a significant weight on the
$(2b_2)^2 \rightarrow (2b_1)^2$ double excitation, and EOM-CCSD
is known to overestimate the energies of double excitations due
to its inability to relax their orbitals.
\cite{Krylov:2008:eom_cc_review,watts1996eomcc,Neuscamman:2016:var_qmc}
The $4^1A_1$ state, on the other hand, does not have any significant
doubly excited components, and so the coupled cluster result should
be more reliable for it.
Indeed, we find that as we go in order of increasing state specificity,
SA(w1) $\rightarrow$ SA(w2) $\rightarrow$ SS, the CASPT2 and MRCI+Q
results for this state move towards the coupled cluster number,
suggesting that there is a modest accuracy improvement to be had for
this state by achieving SS-CASSCF.
Recalling that overall cost in large systems is dominated by
post-CASSCF methods, we see that accuracy improvements of this
type should be achievable at a negligible extra cost compared
to a SA approach.

\begin{table}[!b]
  \normalsize
  \caption{\label{tab:MgO_state} \normalsize 
  Information for the eight lowest MgO $^1A_1$ states
  in the initial LDA-CASCI calculation, including
  their dipole moments ($\mu$, in Debye),
  dominant excitation characters (DECs) and primary active space configurations (PASCs).
  For states with two important configurations of nearly
  equal weight, we have listed both.
    See Figure \ref{fig:MgO_lda_orbs} in the SI for the LDA orbital shapes.
    Note that we order the states' labels within a category
    (GS for ground state, M for missing, V for valence, CT for charge transfer)
    by their SS-CASSCF energies
    and that the ordering of CT3 and CT4 inverts during SS optimization.
  }
\begin{tabular}{ c c r@{.}l c c }
    \hline \hline		
    State & Label & \multicolumn{2}{c}{$\hspace{2mm}\mu$}  & DECs &PACSs\\
    \hline
    $1^1A_1$ & GS   & --3&95 & N/A                                &$5\sigma^2 2\pi^4 6\sigma^2$      \\
    $2^1A_1$ & M1   & --5&39 & $6\sigma \rightarrow 7\sigma$      &$5\sigma^2 2\pi^4 6\sigma^1 7\sigma^1$                             \\
             &      &  \multicolumn{2}{c}{$\hspace{2mm}$}   & $2\pi \rightarrow 3\pi$            &$5\sigma^2 2\pi^3 6\sigma^2 3\pi^1 $\\
    $3^1A_1$ & V1   & --4&88 & $2\pi \rightarrow 3\pi$            &$5\sigma^2 2\pi^3 6\sigma^2 3\pi^1 $   \\
    $4^1A_1$ & V2   & --5&93 & $6\sigma \rightarrow 8\sigma$      &$5\sigma^2 2\pi^4 6\sigma^1 8\sigma^1$ \\
    $5^1A_1$ & CT1  &   3&84 & $2\pi^2 \rightarrow  7\sigma^2$    &$5\sigma^2 2\pi^2 6\sigma^2 7\sigma^2 $   \\
    $6^1A_1$ & CT2  &   3&93 & $2\pi^2 \rightarrow  7\sigma^2$    &$5\sigma^2 2\pi^2 6\sigma^2 7\sigma^2$   \\
     &   & \multicolumn{2}{c}{$\hspace{2mm}$} & $6\sigma^2 \rightarrow  7\sigma^2$                  &$5\sigma^2 2\pi^4 7\sigma^2 $   \\
    $7^1A_1$ & CT4  &   2&33 & $2\pi6\sigma \rightarrow 3\pi7\sigma$  &$5\sigma^2 2\pi^3 6\sigma^1 3\pi^1 7\sigma^1$\\
    $8^1A_1$ & CT3  &   3&66 & $2\pi6\sigma \rightarrow 3\pi7\sigma$  &$5\sigma^2 2\pi^3 6\sigma^1 3\pi^1   7\sigma^1$\\
      &   & \multicolumn{2}{c}{$\hspace{2mm}$}& $2\pi^2 \rightarrow  7\sigma^2$    &$5\sigma^2 2\pi^2 6\sigma^2 7\sigma^2$   \\
    \hline \hline
  \end{tabular}

\end{table}

\subsection{MgO}
\label{sec:mgo}

The final system we investigate is MgO
at a bond distance of 1.8 \r{A}.
Like LiH, the ground state at this near-equilibrium geometry
has ionic character, but in MgO the closed-shell determinant
that dominates the ground state wave function is somewhat
doubly ionic, with both of Mg's 3s electrons moving
into a bonding $\sigma$ orbital with substantial O  2p
character.
We therefore expect to find a challenging assortment
of low-lying excitations including double charge
transfers that return the system to more neutral
states as well as excitations that retain the ground
state's ionic nature.
Indeed, Table \ref{tab:MgO_state} shows that the eight
lowest-lying $^1A_1$ states in an initial
CASCI calculation in the LDA orbital basis (LDA-CASCI) contain
three excited states whose dipoles suggest that they largely
retain the ground state's ionic nature, as well as
four charge transfer states whose electron densities
have shifted significantly towards the Mg atom.
\begin{figure*} [!t]
  \includegraphics[width=15cm]{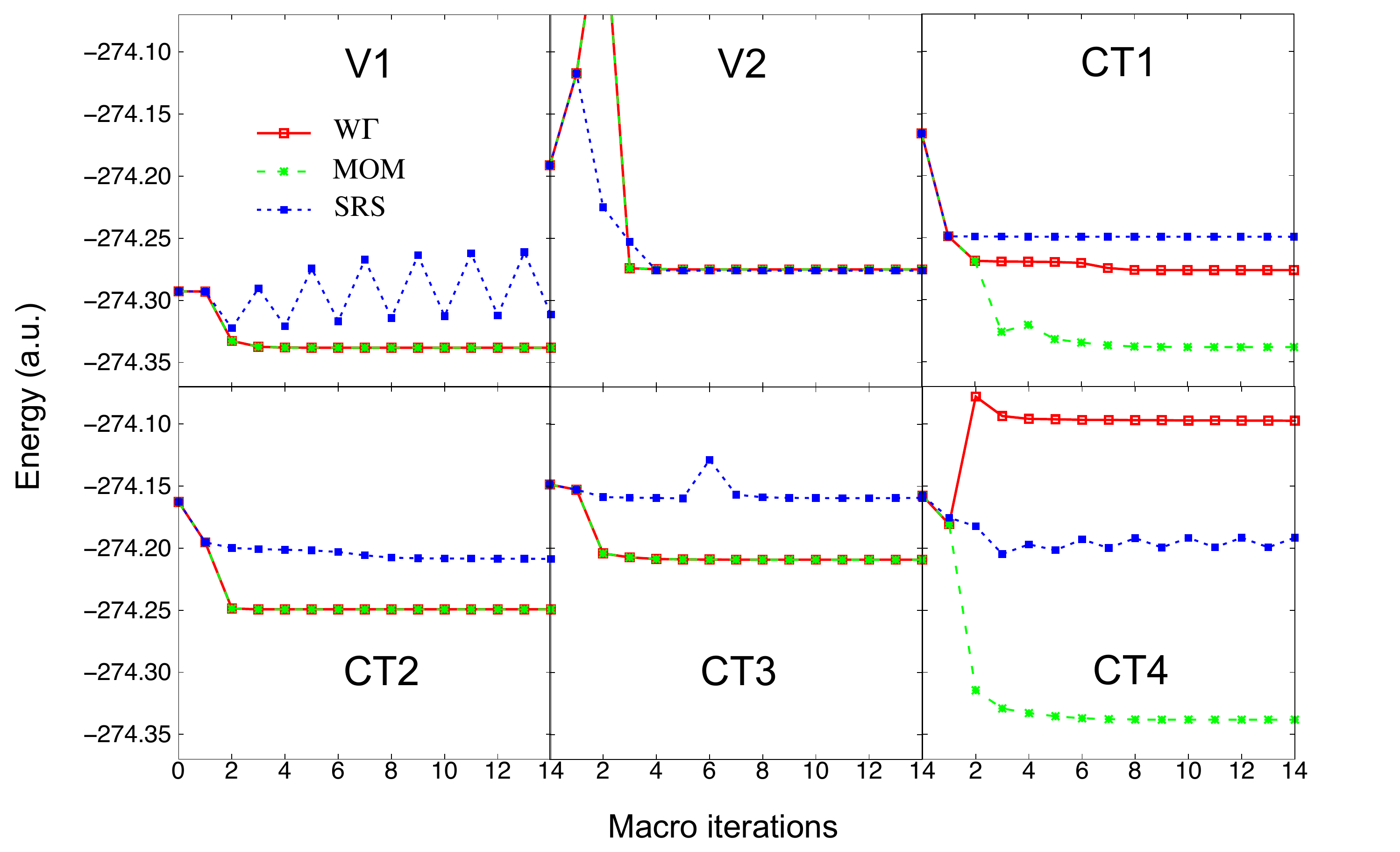}
  \caption{\normalsize
  The selected CASCI root's energy at each macro iteration
  during SS-CASSCF optimizations of six MgO excited
  states under the the SRS, MOM, and $W\Gamma$
  approaches to root selection.
  }
  \label{fig:mgo_iter}
\end{figure*}
The prevalence of double excitations and even double charge
transfers among these states makes MgO a clear example where
a multi-reference treatment is necessary,
but the mixture of neutral and ionic states makes it hard to
know a priori how to construct a SA scheme that treats all
states fairly.
Ideally, this concern could be bypassed via state-specific
optimization, in which the CASSCF energy stationary point
corresponding to each of these states was located.
However, as we now discuss, MgO proves to be especially
difficult in this regard, with SRS failing to converge
to the targeted excited state in every case
and the MOM and $W\Gamma$ approaches succeeding
in only 4 and 6 out of the 7 cases, respectively.

\begin{figure} [!t]
  \includegraphics[width=8.5cm]{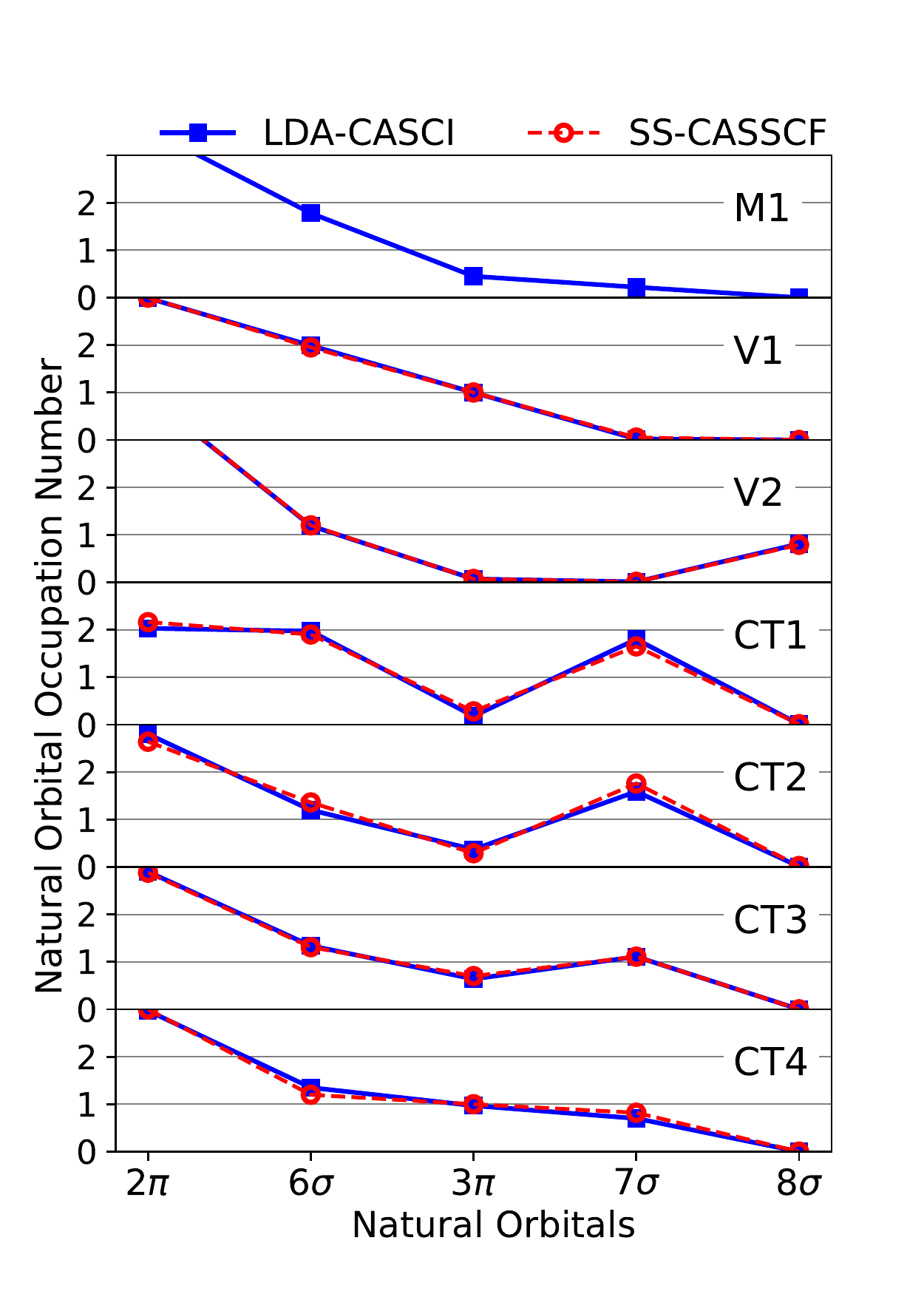}
  \caption{\normalsize
  MgO's excited states may be distinguished using
  natural orbital occupation numbers.
  }
  \label{fig:mgo_occ}
\end{figure}

To begin, we emphasize that the SRS approach fails to converge
to the initially targeted state in every one of the seven
MgO excited states studied here.
That said, Figure \ref{fig:mgo_iter}
shows that although there
are cases where SRS does not converge at all, it often
finds a stationary point corresponding to
a different state than the one originally targeted.
This is true even when targeting state V2, which in
Figure \ref{fig:mgo_iter} looks like a success.
However, inspecting the converged wave function's natural
orbital occupation numbers and comparing to those
in Figure \ref{fig:mgo_occ}
(and Table \ref{tab:MgO_occ} in the SI),
we find that when targeting state V2, SRS
converges to the nearly-degenerate stationary
point belonging to state CT1.
Similarly, when targeting CT1, CT2, and CT3,
SRS converges to the stationary points for CT2, CT3, and a
state outside our set of seven, respectively.

Compared to SRS, both the MOM and $W\Gamma$ approaches 
prove substantially more effective.
For the six states shown in Figure \ref{fig:mgo_iter},
they at least both converge to a stationary point
in every case.
Careful inspection of the final wave functions,
including the natural orbital occupation analysis of
Figure \ref{fig:mgo_occ}, shows that $W\Gamma$ in fact
converges to the intended stationary point in all six
cases.
In contrast, MOM succeeds in only four cases: when
attempting to find the stationary point for the CT1 and
CT4 states, it collapses to the V1 stationary point.
We therefore see that, as in other systems in this study,
the $W\Gamma$ approach proves more effective for
SS-CASSCF than the CI-vector-based MOM, while both
of these greatly outperform SRS.
That said, we note that none of these methods succeeded
in converging to the M1 stationary point, and indeed we
have not been able to locate this point by any means.
Thus, while the $W\Gamma$ approach is promising, it
would clearly benefit from further improvements in
future work.

To investigate whether these SS-CASSCF solutions
offer advantages when using post-CASSCF methods,
we must first decide on SA schemes to compare them
against.
There are of course an infinite number
of possible SA weighting schemes, and it can be
difficult to predict a priori what will be most
effective.
While this difficulty is especially concerning
in a case like MgO where different states have
significantly different charge distributions,
we have used two SA schemes here that each try
to achieve simple forms of balance.
First, we take an equally-weighted
8-state SA approach, which we denote as SA(8).
Second, we attempt to minimize the number of
states involved in the average by only including
enough states such that the desired state is
present, and equally weight the states in the
average.
This SA($N$) approach is much less straightforward,
because the $N$th initial LDA-CASCI root
does not necessarily show up in the same order
(or at all) in an $N$-state SA.
States V1, CT1, CT2, and CT3, show up as the $N$th
root in equal-weight $N$-state SA calculations for
$N=$2, 3, 5, and 8, respectively.
However, the near-degeneracy between V2 and CT1
at the CASSCF level prevents V2 from showing up
at all in a 4-state SA.
The minimum number of states to include in order
for V2 to show up turned out to be $N=6$,
in which case it appears as the 6th root.
State CT4 was even more problematic, and indeed
does not show up at all in equal-weight SA
calculations for any $N\leq 8$, which is
unsurprising in light of how much higher in
energy its stationary point in Figure
\ref{fig:mgo_iter} is relative to the initial
LDA-CASCI energy.
Thus, in summary, CT4 was not included in the SA
comparison, while the SA($N$) approach for
states V1, V2, CT1, CT2, and CT3 uses
the values $N=2$, 6, 3, 5, and 8, respectively.

\begin{table}[!b]
  \normalsize
  \caption{\label{tab:MgO_gap} \normalsize Excitation energies for six $^1A_1$ excited states of MgO.
  CASPT2 used an IPEA level shift of 0.2 a.u.\ to
  avoid intruder states.
  In some cases, states were not found (NF) in the SA approach,
  and in others, the MRCI+Q Davidson solver did
  not converge (NC).
  }
  \begin{tabular}{ccccccccc}
    \hline \hline		
    Method	      &
    &V1 &V2 &CT1 &CT2 &CT3 &CT4 \\
    \hline 
    LDA-CASCI      &
    &3.70     &6.46      &5.10      &7.14   &8.16  &8.07      \\	
    LDA-CASPT2  &
    &3.80     &6.42      &5.68      &6.21   &6.02  &7.31       \\
    LDA-MRCI+Q  &
    &3.81     &6.33      &5.92      &5.59   &6.57 & NC \\
    \hline 
    SA(8)-CASSCF  &
    &3.97     &7.24      &5.10      &5.67   &6.94& NF \\	
    SA(8)-CASPT2  &
    &3.91     &6.76      &5.68      &6.23   &7.02& NF \\
    SA(8)-MRCI+Q  &
    &3.81     &6.31      &5.92      &6.46   &6.66& NF \\
    \hline
    SA($N$)-CASSCF&
    &4.10     &6.76      &5.29      &5.94   &6.94& NF \\
    SA($N$)-CASPT2&
    &3.89     &6.57      &5.66      &6.32   &7.02& NF \\
    SA($N$)-MRCI+Q&
    &3.76     &6.39      &5.99      &6.46   &6.66& NF \\ 
    \hline
    SS-CASSCF	  &
    &4.88     &6.60      &6.57     &7.16   &8.39&11.47     \\
    SS-CASPT2	  &
    &3.76     &6.55      &5.73     &6.32   &6.91&10.44     \\
    SS-MRCI+Q	  &
    &3.79     &6.52      &5.97     &6.44   &6.76& NC \\
    \hline \hline
  \end{tabular}
\end{table}

\begin{figure*} [!t]\centering
  \includegraphics[width=12.0cm]{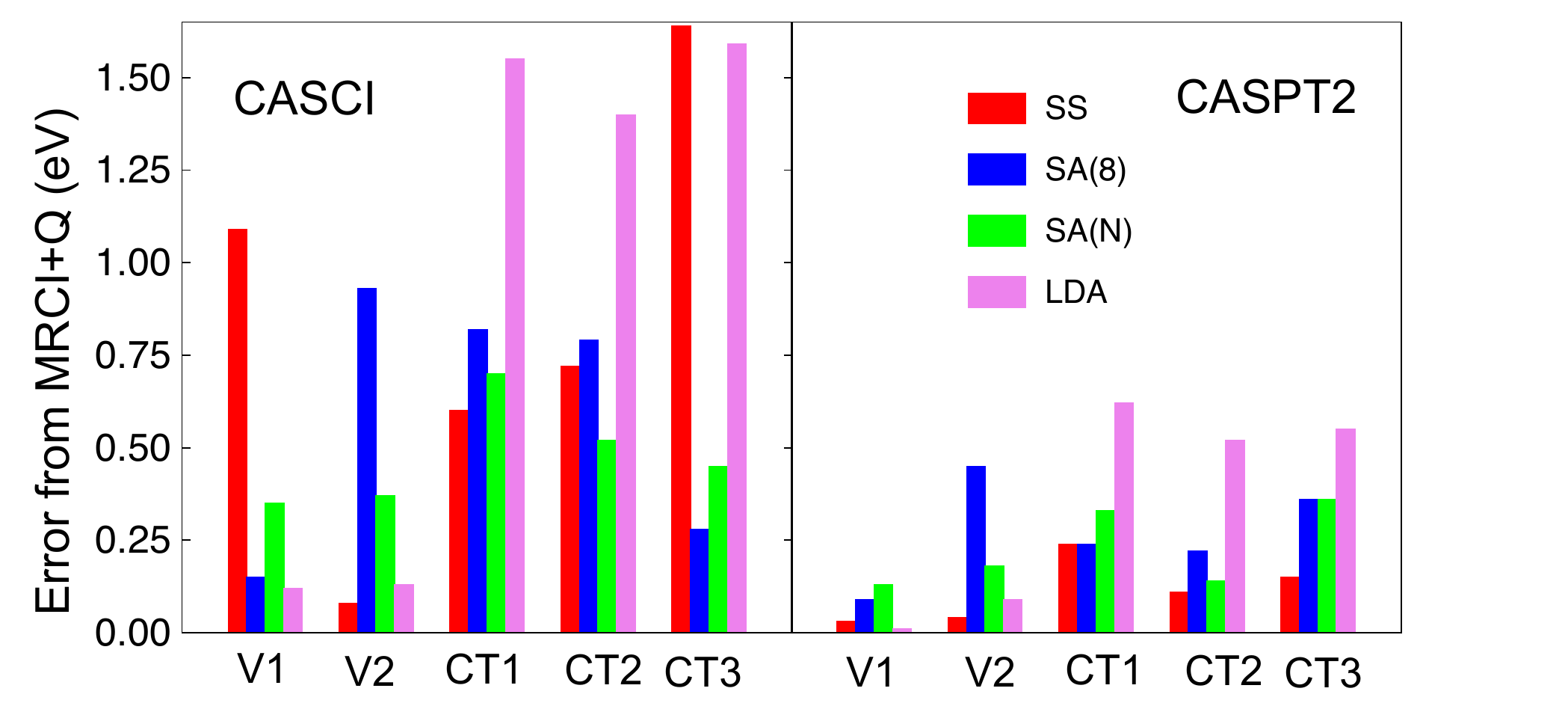}
  \caption{\normalsize
  For different sets of molecular orbitals
  (SS-CASSCF, SA(8)-CASSCF, SA($N$)-CASSCF, and LDA)
  we compare CASCI and CASPT2 excitation energies
  with MRCI+Q.
  In each case, all three methods are carried out in the same
  basis and the difference taken between the
  excitation energy for the method in
  question (CASCI on the left panel and CASPT2 on the right)
  and the excitation energy for MRCI+Q.
  Note that CASPT2 used an IPEA level shift of 0.2 a.u.\ to
  avoid intruder states.
  }
  \label{fig:MgO_err}
\end{figure*}

The LDA-CASCI, CASSCF, CASPT2, and MRCI+Q excitation
energies for the six $^1A_1$ excited states for which
SS-CASSCF stationary points were found are shown in
Table \ref{tab:MgO_gap}.
In Figure \ref{fig:MgO_err}, we show the difference
from the corresponding MRCI+Q excitation energies
for both CASSCF and CASPT2 when working in different
methods' (LDA, SA(8), SA($N$), SS-CASSCF) orbital
bases.
In every case, the SS-CASSCF orbital basis leads to
the smallest difference between CASPT2 and MRCI+Q
excitation energies, which supports the hypothesis
that state-specific orbital relaxations should be
beneficial when relying on the perturbative
assumption that the CASSCF wave function is close
to the exact wave function.
Unsurprisingly, working in the LDA orbital basis
led to the largest differences, while the SA
approaches were in between these two extremes.
While SS-CASSCF thus appears to offer improvements
for CASPT2, it is important to note that its CASSCF
excitation energies are often not closer to the
corresponding MRCI+Q when compared to the situation
in SA-CASSCF, as revealed by the left panel of
Figure \ref{fig:MgO_err}.
This result should not be surprising, as CASSCF
lacks all out-of-active-space weak correlation
effects, the size of which is expected to differ
significantly for different states.
For example, neutral states have roughly 12 and
8 electrons located on the Mg and O atoms,
respectively, while the doubly ionic ground
state has 10 and 10, and so these states
have different numbers of electrons in close
proximity to each other.
As most of the energetic effects coming from
weak electron correlations are local in nature,
a crude accounting of how many local electron
pairs can be enumerated on each atom
(12 choose 2 and 8 choose 2
compared to twice 10 choose 2)
suggests that the size of weak correlation
effects should be different for different states.
Thus, while SS-CASSCF does not and is not expected
to bring the CASSCF energetics closer to those of
MRCI+Q, it does serve as a better platform for
CASPT2 than either of the two SA approaches.



\section{Conclusion}
\label{sec:conclusion}

We have presented an approach for locating state-specific CASSCF energy stationary
points that overcomes root flipping via a metric of quality based
on an excited state variational principle and density matrix differences.
This approach was inspired by both the maximum overlap method and recent
progress in excited state variational methods, as well as the twin goals
of controlling cost and maintaining compatibility with widely used tick-tock
methods that treat orbital and CI coefficients in separate optimization steps.
Our central finding is that this approach is highly effective in the face of
both root flipping and near-degeneracies, significantly outperforming both
simple root selection and a CI-vector-based adaptation of the maximum overlap method.
Although the improvements can be modest, we find that, compared to state averaging,
state-specific CASSCF provides a superior starting point for CASPT2 in that it brings
this method's excitation energies more closely in line with those of the more reliable
but expensive MRCI+Q.
As the additional cost of converging each CASSCF state individually is typically
small compared to the cost of post-CASSCF methods, we are happy to recommend that
applications of this approach be explored across a wider spectrum of chemical systems.

Aside from post-CASSCF accuracy improvements, the ability to locate individual
excited states' energy stationary points delivers a number of important and useful properties.
Unlike state averaged CASSCF, the excited state specific version retains the
size extensivity and size consistency of its ground state counterpart, which
implies that its CASSCF excitation energies will be size intensive.
Furthermore, stationarity with respect to the energy makes the calculation of many
properties, not least of which are the nuclear gradients, significantly more
straightforward as it avoids the need for Lagrangian or Z-vector techniques.
Finally, state-specific optimization avoids the need to decide on how many states
to average and what weights to use, choices that are both difficult to justify
a priori and which can have significant effects on CASSCF and post-CASSCF energetics.

While we have presented evidence in this study of how our approach
to stationary points can be beneficial to CASPT2, other excited state methods
are likely to benefit as well.
For example, the most difficult optimization stage in the application of
variational Monte Carlo to excited states tends to be the step in which
orbital relaxations are enabled.
A question that we are very eager to answer in future work is thus whether
orbitals optimized for an individual excited state by CASSCF could serve
as a good alternative to those optimized by Monte Carlo, and whether
these two sources of optimized orbitals make any difference in practice
when fed in to the diffusion Monte Carlo method.
Back in the area of quantum chemistry, it is quite possible that the
recently-introduced excited state mean field method could be
accelerated by tick-tock optimization methods, at which point it too would
come face to face with root flipping issues that the $W\Gamma$ approach
presented here might alleviate.
Finally, it will be exciting to pair this improved state-specific
methodology with large active space methods, which already offer substantial
advantages when dealing with root flipping.

Looking forward, there are numerous promising ways to extend and apply
this new methodology.
Although we have chosen to avoid the double excitation components of the
variational principle in the present study to keep the approach simple
and inexpensive, an excellent approximation to the effect of this term
should be accessible via the application of modern selective CI integral
handling technology.
Similarly, although the two-body reduced density matrix in the full
orbital space may be impractical to employ for root comparisons
due to high memory demands, one can imagine inspecting it in
a second, much larger active space.
As we have already identified at least one state in MgO that
continues to defy all of the attempted state specific optimizations,
we are eager to explore these extensions in order to make the methodology
more reliable and robust.
In terms of applications, the freedom to define the target state in
whichever way the user chooses would appear to be a natural fit
when trying to optimize the two (or more!) adiabatic states
near a conical intersection.
While these states mix freely at the intersection itself and so
differentiating between them in terms of their diabatic character
is less meaningful, away from but nearby the intersection
the use of diabatic states as the targets for state-specific
CASSCF optimization could help identify which adiabatic states
are which and also help ensure that all the states in question
were found in the state specific optimization process.
More generally, it will be interesting to investigate how a more
aggressively state-specific approach to excited states in CASSCF
can be applied to the ongoing challenge of handling the strong
excited state orbital relaxations common to core and charge
transfer excitations.

\section*{acknowledgement}
This work was supported by
the Early Career Research Program
of the Office of Science, Office of Basic Energy Sciences,
the U.S. Department of Energy, grant No.\ {DE-SC0017869}.
JARS acknowledges additional support from
the National Science Foundation's
Graduate Research Fellowships Program.
Calculations were performed both on our own desktop computers
and using the UC Berkeley Savio computer cluster.

\bibliography{main}


\clearpage
%

\section*{Supporting Information}

\section*{Tracking excited states in wave function optimization using \\ density matrices and variational principles} 

{\it Lan Nguyen Tran, Jacqueline A. R. Shea and Eric Neuscamman}

\beginsupplement

\begin{table}[!h]
  \normalsize
  \caption{\label{tab:lih_ex} \normalsize Excitation energies (in eV) for $A^1\Sigma^+$ of LiH from FCI, SA- and SS-CASSCF at different bond lengths (in \r{A}).}
  \begin{tabular}{cccccccccc}
    \hline \hline		
    &R  &FCI	&SA-CASSCF &SS-CASSCF \\
    \hline
    &1.2 &3.85	&3.47	&3.58 \\
    &1.4 &3.71	&3.36	&3.47 \\
    &1.6 &3.47	&3.14	&3.26 \\
    &1.8 &3.18	&2.88	&3.01 \\
    &2.0 &2.86	&2.61	&2.73 \\
    &2.2 &2.55	&2.34	&2.45 \\
    &2.4 &2.25	&2.09	&2.18 \\
    &2.6 &1.98	&1.87	&1.94 \\
    &2.8 &1.75	&1.68	&1.74 \\
    &3.0 &1.56	&1.54	&1.58 \\
    &3.4 &1.36	&1.39	&1.41 \\
    &3.8 &1.35	&1.41	&1.42 \\
    &4.2 &1.45	&1.51	&1.51 \\
    \hline \hline
  \end{tabular}
\end{table}

\begin{table*}[!h]
  \normalsize
  \caption{\label{tab:MgO_occ} \normalsize Occupation numbers and dipole moments $\mu$ (in Debye) of eight lowest $^1A_1$ states including the ground state from LDA-CASCI and SS-CASSCF calculations for MgO.}
  \begin{tabular}{ccccccccccc}
    \hline \hline		
    &CASCI &\multirow{2}{*}{Label}    &\multirow{2}{*}{Method}	 &\multicolumn{6}{c}{Occupation numbers} &\multirow{2}{*}{$\mu$}\\
    \cline{5-10}
    &ordering & & &$5\sigma$&$2\pi$ &$6\sigma$ &$3\pi$ &$7\sigma$ &$8\sigma$ &\\
    \hline	
    &$1^1A_1$ &GS &CASCI &2.00     &3.92 &1.63 &0.07 &0.36 &0.00 &--3.95\\
    &         &&SS-CASSCF &2.00&3.93 &1.73 &0.07 &0.27 &0.01&--4.62 \\
    &$2^1A_1$ &M1 &CASCI &2.00    &3.55 &1.78 &0.45 &0.22 &0.00&--5.39\\
    &$3^1A_1$ &V1 &CASCI &2.00     &2.99 &1.99 &1.00 &0.02 &0.00&--4.88 \\
    &         &&SS-CASSCF &2.00 &3.00 &1.95 &1.00 &0.05 &0.00&--4.58 \\
    &$4^1A_1$ &V2 &CASCI    &2.00 &3.93 &1.19 &0.07 &0.01 &0.81&--5.93\\
    &         &&SS-CASSCF &2.00&3.93 &1.20 &0.07 &0.01 &0.79&--5.48 \\
    &$5^1A_1$ &CT1&CASCI    &2.00 &2.03 &1.97 &0.18 &1.80 &0.01&3.84\\
    &         &&SS-CASSCF &2.00&2.16 &1.90 &0.28 &1.65 &0.01&3.14 \\
    &$6^1A_1$ &CT2&CASCI   &2.00  &2.80 &1.20 &0.37 &1.59 &0.01&3.93\\
    &         &&SS-CASSCF &2.00&2.64 &1.36 &0.29 &1.76 &0.02&3.48 \\
    &$7^1A_1$ &CT4&CASCI  &2.00   &2.97 &1.35 &0.97 &0.70 &0.00&3.66\\
    &         &&SS-CASSCF &2.00&3.00 &1.20 &1.00 &0.82 &0.00&3.16 \\
    &$8^1A_1$ &CT3&CASCI  &2.00   &2.90 &1.34 &0.64 &1.11 &0.00&2.33 \\
    &         &&SS-CASSCF &2.00&2.88 &1.31 &0.70 &1.11 &0.00&2.18 \\ 
    \hline \hline
  \end{tabular}
\end{table*}

\begin{figure*} [!t]\centering
  \includegraphics[width=14.0cm]{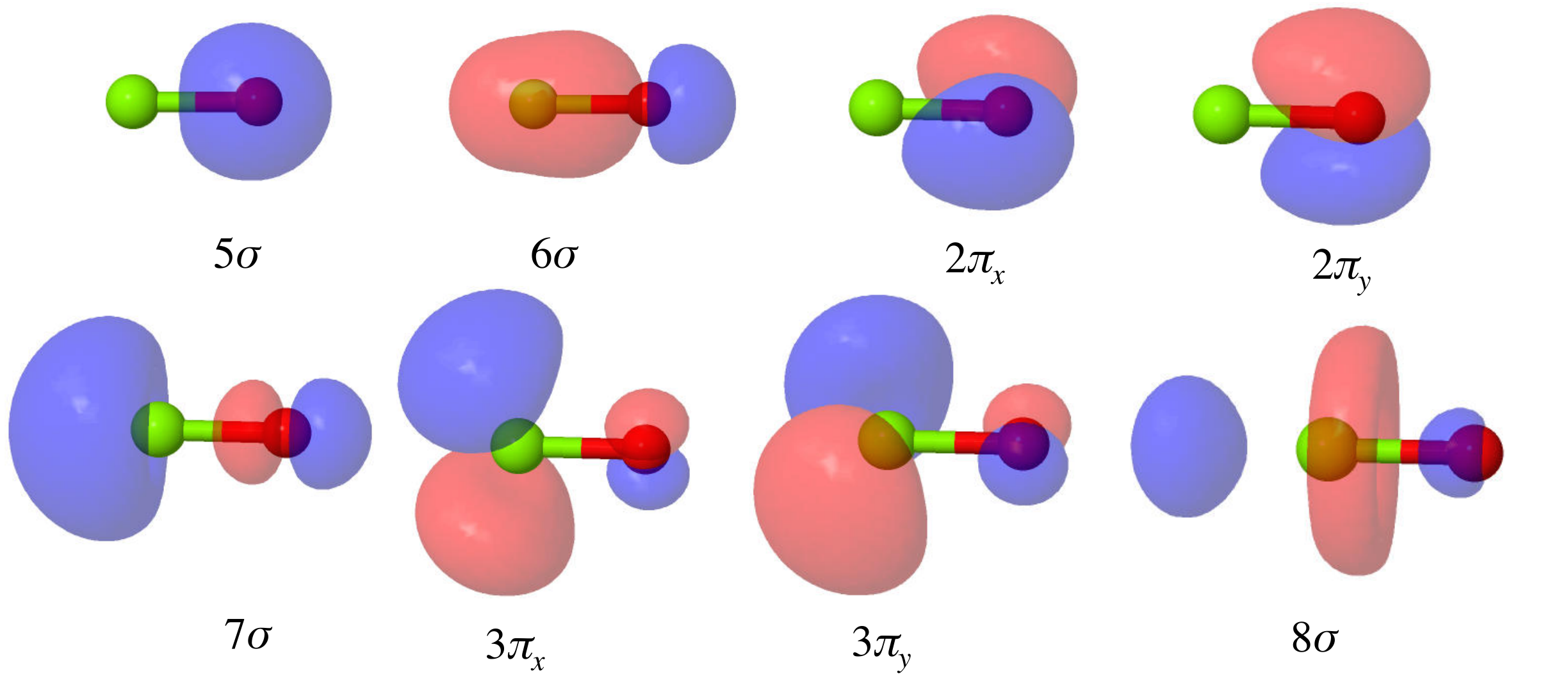}
  \caption{\normalsize
    Shapes of the eight LDA orbitals used to
    construct the active space for MgO.
    In each image, the Mg atom is on the left.
  }
  \label{fig:MgO_lda_orbs}
\end{figure*}

\end{document}